# QUANTUM MECHANICAL DISCLOSURE OF THE CLASSICAL ADIABATIC CONSTANCY OF PV$^\gamma$ FOR AN IDEAL GAS, AND FOR A PHOTON GAS

Metin Arik (metin.arik@boun.edu.tr), Bogaziçi University, Istanbul, Turkey
Tolga Yarman, Department of Engineering, Okan University, Istanbul, Turkey & Savronik, Organize Sanayii Bölgesi, Eskisehir, Turkey, tolgayarman@gmail.com
Alexander L. Kholmetskii[*], Department of Physics, Belarus State University, Minsk, Belarus, khol123@yahoo.com

**Abstract**
Previously, we established a connection between the macroscopic classical laws of gases and the quantum mechanical description of molecules of an ideal gas (T. Yarman *et al*. arXiv:0805.4494). In such a gas, the motion of each molecule can be considered independently on all other molecules, and thus the macroscopic parameters of the ideal gas, like pressure *P* and temperature *T*, can be introduced as a result of simple averaging over all individual motions of the molecules. It was shown that for an ideal gas enclosed in a macroscopic cubic box of volume *V*, the *constant,* arising along with the *classical law of adiabatic expansion, i.e. PV$^{5/3}$=constant*, can be explicitly derived based on quantum mechanics, so that the constant comes to be proportional to $h^2/m$; here *h* is the Planck Constant, and *m* is the relativistic mass of the molecule the gas is made of. In this article we show that the same holds for a photon gas, although the related setup is quite different than the previous ideal gas setup. At any rate, we come out with $PV^{4/3} \sim hc = Constant$, where *c* is the speed of light. No matter what the dimensions of the *constants* in question are different from each other, they are still rooted to universal constants, more specifically to $h^2$ and to *hc,* respectively; their ratio, i.e. $V^{1/3} \sim h/mc$, interestingly pointing to the de *Broglie relationship's* cast.

## 1. Introduction

It is known that the question of finding a connection between the Boltzmann constant *k* and the Planck constant *h* remains unanswered. In the previous work [1] we have shown that such an effort is in vain, for as we have elaborated on, one can only define one of these quantities, based on the other. Instead though, we established an *organic bridge* between the *macroscopic classical laws of gases* and the *quantum mechanical description of molecules* of an *ideal gas,* within the framework of a gas relationship involving neither *k* nor *h.* Along this line, it would be fair to recall that in particular, de Broglie already in his doctorate thesis has brilliantly applied his relationship (associating a wave length with the momentum of a moving particle) to the statistical equilibrium of gases [2], but did not advance his idea, to see whether one can, along such a line, obtain anything related to the laws of gases, established long ago, in 1650. Modern statistical physics, despite huge efforts to draw a *parallelism* between the classical law of gases and quantum mechanics, does not yet appear at the level of directly implementing the two disciplines in question, into each other, the way we did in ref. [1].

In an ideal gas, by definition, one proposes to consider the motion of each molecule independently on all other molecules. Accordingly the macroscopic parameters of the ideal gas, such as pressure *P* and temperature *T*, can be introduced as a result of simple averaging over all individual motions of molecules. In the mentioned work [1] we had thus shown that for an ideal gas enclosed in a macroscopic cubic box of volume *V*, the classical law of adiabatic expansion,

$$PV^\gamma = constant, \qquad (1)$$

can be derived based on simple quantum mechanics. A principal advantage of such a quantum mechanical analysis is the explicit determination of the constant of eq. (1), which turns out to be $h^2 n^2/(4m_0)$ for a gas made of just one molecule of mass $m_0$. Here *n* is the integer number characterizing the energy level the molecule in the simplifying assumption, where all three quantum numbers $n_x$, $n_y$, and $n_z$ are equal to $n$, thus equal to each other. The result can easily be extended and, via averaging, generalized to a given set of molecules.

---
[*] The corresponding author.

Below, we first summarize the previous work [1], which constitutes the basis of the present contribution (section 2). Then we undertake the case of a photon gas (section 3). We show that the $PV^\gamma$=*constant* holds for a photon gas, too, with $\gamma$=4/3. The constant coming into play being still nailed to the Planck constant $h$. Finally a conclusion is drawn in section 4.

## 2. The harmony of the phenomenological laws of gases with quantum mechanics based on the constancy of $PV^\gamma$ for an adiabatic transformation

As anticipated previously [1], the relationship (1) for an adiabatic transformation, an ideal gas displays, constitutes an efficient *check point* of the *compatibility* of the *macroscopic laws of gases* and *quantum mechanics*. Below, for simplicity, we will operate with one mole of gas. We could well operate with just a single molecule, and the results would still be the same, since in an ideal gas the molecules are supposed not to interact with each other.

Thus, the second author *et al.* have previously proposed to calculate specifically, the constant in question, within a quantum mechanical framework. Below we summarize the derivation.

Eq. (1) involves the usual definition

$$\gamma = C_P/C_V , \qquad (2)$$

where

$$C_V = \frac{3}{2}R, \quad C_P = \frac{5}{2}R, \qquad (3), (4)$$

$C_V$ being the heat to be delivered to one mole of ideal gas at *constant volume* to increase the temperature of the gas as much as 1° K, and $C_P$ being the heat to be delivered to one mole of ideal gas at *constant pressure* to increase its temperature, still as much as 1° K, and $R$ is the gas constant. Eqs. (3) and (4) are exact, when internal energy levels of molecules are not excited. By definition, such an approximation is fulfilled for an ideal gas. Hence we have

$$\gamma = 5/3 . \qquad (5)$$

It is worth to emphasize that eq. (2) would remain valid, even if the ideal gas consists of a single molecule. It may indeed be recalled that, within the frame of the kinetic theory of gases, one first expresses the pressure for just one molecule of gas, before he proceeds for very many more, making up the gas of concern. This is how, one formulates the macroscopic pressure, the gas exerts on the walls of its container. In what follows, we determine the constant of eq. (1) first for slowly moving molecules of the ideal gas (sub-section 2.1), and then for relativistically moving molecules (sub-section 2.2)

### 2.1. The non-relativistic case

Let us consider a non-relativistic particle of rest mass $m_0$ at a fixed internal energy state, located in a macroscopic cube of side $L$. The non-relativistic Schrödinger equation furnishes the $n^{th}$ energy $E_n$ is

$$E_n = \frac{h^2}{8m_0}\left(\frac{n_x^2}{L^2} + \frac{n_y^2}{L^2} + \frac{n_z^2}{L^2}\right) = \frac{h^2(n_x^2 + n_y^2 + n_z^2)}{8m_0 L^2}, \qquad (6)$$

where we denoted $n_x$=1,2,3…$n$, $n_y$=1,2,3…$n$, $n_z$=1,2,3…$n$ the quantum numbers to be associated with the corresponding wave function dependencies on the respective directions $x$, $y$ and $z$. Hereinafter, for brevity, while writing $E_n$, we introduced the subscript "$n$" to denote the given state characterized by the integer numbers $n_x$, $n_y$ and $n_z$, so each "$n$" in fact, represents a set of three integer numbers.

For an ideal gas confined in an infinitely high box, the potential energy input to the Schrödinger equation is null everywhere inside the box. (It is evidently infinite at the borders). Hence for a non-relativistic particle, we have

$$E_n = \frac{1}{2} m_0 v_n^2, \tag{7}$$

$v_n$ being the velocity of the particle at the $n^{th}$ energy level.

At the given energy level, the pressure $p_n$ exerted by the single particle on the walls, after averaging over three dimensions, becomes [1]

$$p_n = \frac{m_0 v_n^2}{3L^3} = \frac{2}{3} \frac{E_n}{L^3}. \tag{8}$$

Now let us calculate the product $p_n V^\gamma$:

$$p_n V^\gamma = \frac{2}{3} \frac{\frac{h^2 \left(n_x^2 + n_y^2 + n_z^2\right)}{8 m_0 L^2}}{L^3} \left(L^3\right)^{5/3} = \frac{h^2 \left(n_x^2 + n_y^2 + n_z^2\right)}{12 m_0}. \tag{9}$$

We observe that the rhs of eq. (9) turns out to be a constant for the given discrete energy level $n$ (specified by the set of $n_x$, $n_y$ and $n_z$) of the particle of mass $m_0$. Recall that the total quantized energy $E_n$ in eq. (6) ultimately determines the quantized velocity $v_n$ of eq. (7) along with its three quantized components.

When it is question of many particles instead of just one, we have to consider the particles at different, possible, quantized states. We can anyway visualize the *average particle* at the $\bar{n}^{th}$ level, thus corresponding to the given temperature of the gas[†] at the given state, and suppose that all other particles behave the same. Furthermore, all three components of the average velocity are expected to be the same in equilibrium state. Thus, we can rewrite eq. (9) for the macroscopic pressure $P_{\bar{n}}$ exerted at the given average state $\bar{n}$ by one mole of gas on the walls of the container:[‡]

$$P_{\bar{n}} V^\gamma = N_A \frac{h^2 \bar{n}^2}{4 m_0}, \tag{10}$$

where $N_A$ is the Avagadro number.

Thus eq. (10) discloses the constant involved by the adiabatic transformation relationship, i.e. eq. (1). Note that at the average state $\bar{n}$ (i.e. at the given temperature), the mean square speed of the gas molecules is $v_{\bar{n}}^2 = \overline{v_n^2}$; the average energy $E_{\bar{n}} = \overline{E_n}$ is furnished accordingly, via the framework of eq.(7).

Thus, we arrive to conclude that the constancy $P_{\bar{n}} V^\gamma$, drawn by an adiabatic transformation of an ideal gas, is nothing but a *macroscopic manifestation of its quantum mechanical behavior*.

The above results, i.e. eq. (8) and eq. (9) would not change, if we operated in, not three dimensions, but just one dimension. The reason is simply that, the factor 1/3 introduced at the level of eq. (8) due to the exercise of three dimensions, would be cancelled out by the factor 3, that would come into play, at the level of eq. (9) due to the introduction of the corresponding three quantum numbers (equal to each other regarding the average energy level we visualized).

---

[†] Note that through an adiabatic transformation of particles in a box, the *"temperature"* will get changed, whereas the quantum denominations associated with the energy levels of these particles, will remain the same; that is the quantum numbers coming into pay, would not get altered. Thus, we have to precise what we mean here, by "temperature". We mean, the "average energy of the constituents in the box, at the given state, prior to the transformation".

[‡] Rigorously speaking, one must write [1] $P_{\bar{n}} V^\gamma = \sum_{i=1}^{N_A} \frac{h^2 \left(n_{ix}^2 + n_{iy}^2 + n_{iz}^2\right)}{12m} = N_A \frac{h^2 \bar{n}^2}{4m}$ along with the definition $\frac{1}{3N_A} \sum_{i=1}^{N_A} \left(n_{ix}^2 + n_{iy}^2 + n_{iz}^2\right) = \bar{n}^2$. Hence it becomes clear that, if all particles bare the same *set of quantum numbers*, each with equal quantum numbers along all three directions, i.e. $n_x = n_y = n_z = \bar{n}$, then $\bar{n}$ becomes $\bar{n} = \sqrt{\overline{n^2}}$.

### 2.2. The relativistic case

We start with a relativistic generalization of eq. (8), which obviously implies the replacement of $m_0$ by the *relativistic mass* of molecule $m=\gamma m_0$:

$$p_n = \frac{mv_n^2}{3L^3} = \frac{m^2 v_n^2}{3mL^3} = \frac{\Lambda_n^2}{3mL^3}, \tag{11}$$

where $\Lambda_n$ being the relativistic momentum to be furnished by de Broglie relationship, so that

$$\Lambda_n = mv_n = \frac{(a\ number\ related\ to\ the\ n^{th}\ quantum\ state) \cdot h}{2L}. \tag{12}$$

Here again to keep a long denomination short, we chose to indicate the state of concern by the mere letter *n*, which in fact should embody the set of three quantum numbers, $n_x$, $n_y$, and $n_z$ each to be associated with the related dimension. What is then *n* that will come to multiply *h*? We do not really have to know it. What we have to know is the square of this number, since in eq. (11) we need $mv_n^2$, and not $mv_n$. Thus what we need, is a *quantum number* to be associated with $v_n^2 = v_x^2 + v_y^2 + v_z^2$, i.e. the square of the velocity. It becomes easy to guess that this number will be

$$n^2 = n_x^2 + n_y^2 + n_z^2. \tag{13}$$

Let us now write eq. (11) for the *average state*:

$$p_n = \frac{\Lambda_n^2}{3mL^3} = \frac{h^2\left(\overline{n_x^2} + \overline{n_y^2} + \overline{n_z^2}\right)}{(3mL^3)(4L^2)} = \frac{h^2 \overline{n}^2}{4mL^5}, \tag{14}$$

where the last equality, as before, implies the equality of the three quantum numbers for the average state we characterize by $\overline{n}$. We can then compose $P_{\overline{n}} V^\gamma$ for one mole gas [see, eq.(9)]:[§]

$$P_{\overline{n}} V^\gamma = N_A \frac{h^2 \overline{n}^2}{4mL^5} (L^3)^{5/3} = N_A \frac{h^2 \overline{n}^2}{4m}. \tag{15}$$

Thence, we come to the conclusion that in the relativistic case, $P_{\overline{n}} V^\gamma$ does not remain constant, since the relativistic mass *m* is not a constant *(albeit the rhs appears to be a constant)*. But $mP_{\overline{n}} V^\gamma$ well is:

$$mP_{\overline{n}} V^\gamma = \frac{\Gamma_{\overline{n}} P_{\overline{n}} V^\gamma}{c^2} = N_A \frac{h^2 \overline{n}^2}{4}. \tag{16}$$

for an adiabatic transformation.

Thus, we have found that in a gas where molecules would bare the relativistic energy $\Gamma_{\overline{n}} = mc^2$, in the average, at the $\overline{n}^{th}$ state, the product $\Gamma_{\overline{n}} P_{\overline{n}} V^\gamma$ through an adiabatic transformation remains constant, which, for one mole of gas, becomes $N_A h^2 \overline{n}^2 c^2 / 4$. What is more, the *adiabatic constancy* of the product $\Gamma_{\overline{n}} P_{\overline{n}} V^\gamma$ is nothing but a macroscopic *quantum mechanical manifestation*, the gas delineates.

---

[§] Note that the total relativistic energy $\Gamma_{\overline{n}} = mc^2$ for the *average particle* is given as

$$\Gamma_{\overline{n}} = mc^2 = \sqrt{m_0^2 c^4 + \Lambda^2 c^2} = \sqrt{m_0^2 c^4 + c^2 h^2 \overline{n}^2 / 4L^2},$$

where we have used eq. (12), along with $\overline{n}^2 = n_x^2 + n_y^2 + n_z^2$, for the *average particle*, which bares three equal quantum numbers; eq. (15) accordingly becomes $P_{\overline{n}} V^\gamma m_0 \sqrt{1 + h^2 \overline{n}^2 / 4m_0^2 c^2 L^2} = N_A \frac{h^2 \overline{n}^2}{4}$.

Note that up to this point, the exponent $\gamma$ in all expressions of (pressure × volume)$^{\gamma}$ was 5/3. It is different for the photon gas, analyzed in section 3, below, though it will still indicate the ratio of specific heats at respectively constant pressure and constant volume.

## 3. Adiabatic transformation of the photon gas

The basic finding we proposed to provide in this section is whether or not a photon gas would fulfill our disclosure about the adiabatic constancy of (pressure × volume)$^{\gamma}$. We will see below that it does. We will accordingly specifically calculate the constant coming into play in *(pressure × volume)$^{\gamma}$=Constant.*

Thus, consider once again a cube of side *L*, with just one photon moving in a perpendicular direction to two surfaces. The total energy *E* of the photon is as usual

$E = \Lambda c$, (17)

where $\Lambda$ is the relativistic momentum of the photon. The force *F* the photon exerts on the wall is

$$F = \frac{\Delta \Lambda}{\Delta t} = \frac{2\Lambda}{2L/c} = \frac{\Lambda c}{L}.$$ (18)

Thus the pressure *p* on the side of the cube, created by the given photon's hits is

$$p = \frac{F}{L^2} = \frac{\Lambda c}{L^3} = \frac{E}{V}.$$ (19)

This is the pressure on the two sides to be perpendicular to the photon direction. However the pressure on the other four sides of the cube, the way we have just set it up, is zero. Thus, were we working in three dimensions, we are to write the *average pressure,* as

$$p = \frac{E}{3V},$$ (20)

or

$E = 3pV$. (21)

A photon, on the other hand, has altogether 3x2=6 degrees of freedom: three components of momentum with two kinds (right handed or left handed) of circular polarization. Thus at the temperature *T*, one can write

$$E = 6\frac{kT}{2} = 3kT,$$ (22)

where *k* is the Boltzmann constant. This makes that for the photon the specific heat $c_V$ at constant volume, becomes

$$c_V = \frac{dE}{dT} = 3k.$$ (23)

In order to derive an expression for $c_p$ (the specific heat at constant pressure), we recall the first law of thermodynamics

$dE = \delta Q - pdV$, (24)

which expresses an increase of the total energy as much as *dE*, if an amount of heat $\delta Q$ is received from the outside, upon which the constituent delivers to the outside the work *pdV*. At the *constant pressure,* the first law of thermodynamics yields:

$$(\delta Q)_p = dE + pdV = 4pdV = \frac{4}{3}dE,$$

where we have used eq.(21). Hence

$$c_p = \left(\frac{\partial Q}{\partial T}\right)_p = \frac{4}{3}\frac{dE}{dT} = 4k,$$ (25)

which, via eq.(23), leads to

$$\gamma = \frac{c_p}{c_v} = \frac{4}{3}.$$ (26)

Let us finally calculate the product $pV^\gamma$ for the photon at hand, taking into account eq. (21):

$$pV^\gamma = pV^{4/3} = \frac{E}{3V}V^{4/3} = \frac{E}{3}L = \frac{h\nu L}{3}, \qquad (27)$$

where $\nu$ is the frequency of the photon of concern, so that $E = h\nu$.

To proceed from here on, we may for simplicity (and without any loss of generality) assume that, the photon moves in one dimension only (and perpendicularly to two parallel sides of the cube of concern). Thus, *classically,* and ultimately *quantum mechanically,* it leads to a standing way inside the box, where the known basic relationship, i.e. $L = \frac{n\lambda_n}{2}$, is to apply, along with $c = \lambda_n \nu_n$, $\lambda_n$ being the wavelength of the photon, and $n$ a corresponding integer number. Hence we finally have

$$p_n V^{4/3} = h\nu_n \frac{n\lambda_n}{6} = \frac{\nu h c}{6}, \qquad (28)$$

where we precise the fact that the pressure $p_n$ is created by the photon being at the $n^{th}$ level in the box.

This relationship is written for just one photon, also for one dimension only. It can easily be generalized to three dimensions and an arbitrary number $N$ of photons confined inside the box, for which one will have to define (just the way we have done for an ordinary gas) an average quantum number $\bar{n}$, to be ascribed to the conveniently defined average energy of the photons of concern:

$$P_{\bar{n}} V^{4/3} = N \frac{\bar{n} h c}{6}; \qquad (29)$$

here $P_{\bar{n}}$ being the average pressure, and $\bar{n}$ the average quantum number defined in the same way, as above. We can see that the product $P_{\bar{n}} V^\gamma$ remains constant, when the number of photons at a fixed energy/frequency state inside the box is a given constant (i.e. there is no absorption of photons by the walls of the box).

## 4. Conclusions

We recall the fact that $PV^\gamma$ stays constant through an adiabatic transformation, has been derived practically in all books on thermodynamics (*e.g.*, [7, 8, 9,10,]), yet based on the purely *phenomenological description of the gas*. However, the *value of the constant* delineated by $PV^\gamma$, to our recollection, is something totally missed over almost a century, after quantum mechanics came into play. As far we could see, throughout, no one seems to have even wondered about the possible value of this constant.

In a previous work, we had calculated this constant for an ideal gas, thus succeeding at the same time to establish an organic link between *classical thermodynamics* and *quantum mechanics* [1]. The essence of our approach is to express the energy, entering into eq.(8) for the pressure exerted by just *one molecule,* through the *quantum mechanical energy eigenvalue* relationship, expressed by eq. (6). We are convinced that this is a warranted procedure, even if the molecule is confined inside a macroscopic recipient.

This way, we could derive the value of the constant delineated by the quantity $PV^\gamma$ for the given average discrete energy level $n$ related to the particle, no matter what the number $n$ might be huge for a macroscopic cube. Then the value of this constant is obtained via simple quantum mechanics and ordinary averaging (see, eq. (10)).

An extension of our approach to the case of relativistically moving particles, composing an ideal gas, led to the constancy of the product (total relativistic energy of particle $\times PV^\gamma$). This result seems to be deep, because this product turns out to be a *Lorentz scalar* (see, eq. (16)). However, further discussion of this result will be presented elsewhere.

Herein we have extended our approach to the photon gas. Because now it is question of a photon, the setup is, as expected, different than the one we have established previously. Nevertheless the result, *cast-wise,* is the same. In other words, the quantity $pV^\gamma$ for a photon is i) a constant, and ii) nothing but $\bar{n}hc/6$.

Recall that for an ordinary gas, the exponent $\gamma$ in all expressions of (pressure × volume)$^\gamma$ was 5/3. It became 4/3 for the photon gas, though it still indicates the ratio of specific heats at respectively constant pressure and constant volume.

No matter what the dimensions of the constants in question are different from each other, it seems of course, striking that, they are still rooted to *universal constants,* more specifically to $h^2$ and to $hc$, respectively, their ratio, i.e. $V^{1/3}=h/mc$, interestingly pointing to the *de Broglie relationship's cast.*

Finally the results of the present paper may not lead to new experimental predictions. Nonetheless, we believe that they have a general significance and unexpectedly show that the phenomenological laws for the ideal gas, and for a photon gas, can be interpreted as a macroscopic manifestation of quantum phenomena, which certainly seems quite unexpected chiefly for an ordinary gas.

Any deviation from $PV^\gamma = constant$ must mean that, one then deals with something else than an ideal gas.